\documentclass[
 preprint,
 amsmath,amssymb,
 aps,
 pra,
floatfix,
]{revtex4-2}

\usepackage{amsmath}
\usepackage{graphicx}
\usepackage{dcolumn}
\usepackage{bm}
\begin{document}

\preprint{OSESPA ARXIV}

\title{Optical Shock-Enhanced Self-Photon Acceleration\\}

\author{P. Franke}
\email{pfranke@ur.rochester.edu}
\author{D. Ramsey} \author{T. T. Simpson} \author{D. H. Froula} \author{J. P. Palastro} 
\affiliation{University of Rochester, Laboratory for Laser Energetics, Rochester, New York 14623, USA}%

\date{\today}

\begin{abstract}
Photon accelerators can spectrally broaden laser pulses with high efficiency in moving electron density gradients driven in a rapidly ionizing plasma. When driven by a conventional laser pulse, the group velocity walk-off experienced by the accelerated photons and deterioration of the gradient from diffraction and plasma-refraction limit the extent of spectral broadening. Here we show that a laser pulse with a shaped space-time and transverse intensity profile overcomes these limitations by creating a guiding density profile at a tunable velocity. Self-photon acceleration in this profile leads to dramatic spectral broadening and intensity steepening, forming an optical shock that further enhances the rate of spectral broadening.  In this new regime, multi-octave spectra extending from $400 \, \mbox{nm}-60 \, \mbox{nm}$ wavelengths, which support near-transform limited $<400 \, \mbox{as}$ pulses, are generated over $<100 \, \mu\mbox{m}$ of interaction length.
\end{abstract}

\maketitle
\section{Introduction}
Broadband sources of coherent radiation find utility across diverse scientific disciplines as experimental drivers and diagnostic tools. State-of-the-art supercontinuum sources, which primarily achieve spectral broadening through Kerr-induced self-phase modulation of ultrashort laser pulses in either gas-filled fibers \cite{SCPCF_Review} or self-guided filaments \cite{FilReview}, routinely generate multi-octave spectra in the infrared to ultraviolet wavelength range $(15 \, \mu \mbox{m} - 100 \, \mbox{nm}$) \cite{LongwaveSC,VUVSCExp1,VUVSCExp2,NatureSCExp,UVFilament}. Such sources have thus far been limited to wavelengths $ > 100 \, \mbox{nm} $, due to a lack of dispersion control and strong absorption in the extreme ultraviolet ($ \lambda = 10 \, \mbox{nm} - 100 \, \mbox{nm}$). Extending coherent broadband sources into the extreme ultraviolet would open new wavelength regimes for spectroscopy and increase the achievable spatial and temporal resolution for applications including single-shot spectral interferometry \cite{as_interfer_application}, transient spectroscopy \cite{as_spect_application}, and coherence tomography \cite{XCT_multishot1}. 
\par
Photon acceleration, \emph{i.e.}, the continuous increase in frequency accrued by a photon in a temporally-decreasing refractive index, offers a promising alternative to Kerr-based supercontinuum sources, with the potential to efficiently generate broadband extreme ultraviolet radiation. A photon accelerated in a rapidly ionizing medium has a frequency $\omega$ that evolves in time according to $ d \omega^2 / d t = \partial \omega_p^2/\partial t$, where $\omega_p \propto \sqrt{n_e}$ is the plasma frequency and $n_e$ is the electron density \cite{PAPaper,PABook}. Maximizing the frequency shift requires the photon to propagate within a sharp electron density gradient over an extended distance. An intense laser pulse can photoionize a dense gas triggering an extremely rapid rise in the electron density. In principle, this allows for large frequency shifts over relatively short interaction lengths. Thus far, however, photon acceleration from optical to extreme ultraviolet frequencies has met with limited success \cite{PAExp1,PAExp2,PAExp3,IWAVPA}.
\par 

\begin{figure}[b]
\includegraphics[scale=1.0]{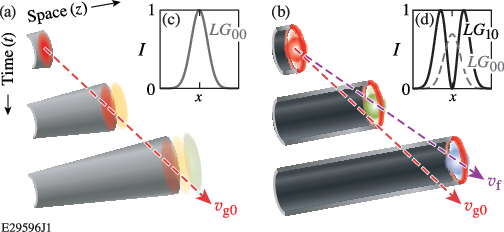}
\caption{\label{fig1} (a) A Gaussian beam drives a (gray) radially convex ionization front at the group velocity of light over approximately a Rayleigh length. Photons diverge from the optical axis due to diffraction and plasma-refraction. Frequency up-shifted photons dephase from the ionization front due to group velocity walk-off, limiting the system to relatively small frequency shifts. (b) A SFF pulse drives a concave ionization front at a tunable focal velocity $v_f \gtrsim v_{g0}$ over a distance much greater than a Rayleigh length. Photons are concentrated near the optical axis and stay in phase with the ionization front, resulting in many photons undergoing a large frequency shift. Transverse intensity profiles of a typical (c) Gaussian pulse and (d) SFF pulse used in 2D simulations.}
\end{figure}
A typical photon accelerator uses a laser pulse focused by conventional optics to drive an axial electron density gradient that moves at the group velocity of the pulse $v_{g} < c$, where $c$ is the vacuum speed of light [Fig 1(a)]. Witness photons injected into the gradient undergo a frequency up-shift, increasing their group velocity. The resulting group velocity mismatch causes witness photons to outrun, or dephase from the gradient, terminating their frequency increase. Diffraction and plasma-refraction cause the drive pulse to diverge from the optical axis, reducing the sharpness of the gradient and the accelerator length, further limiting the maximum frequency shift.
\par
Here we introduce a scheme [Fig. 1(b)], which largely eliminates the adverse effects of diffraction, plasma-refraction and dephasing by combining spatiotemporal \cite{FlyingFocus1,FlyingFocus2,DLWFA,DLWFA2,NLFF,Abouraddy} and transverse intensity profile shaping of the laser pulse \cite{SpiralPP1,OAMReview,OASpiral}. This structured flying focus (SFF) pulse has a shaped spatiotemporal profile that creates a far-field intensity peak propagating at a tunable velocity, $v_f$, over an extended focal range much greater than a Rayleigh length, $Z_R$. This flying focus effect results from different time slices within the pulse focusing to different positions in the far-field. The cylindrically symmetric transverse intensity profile of the SFF increases from a moderate value on the optical axis to a peak at larger radii before falling to zero, so that higher off-axis optical field ionization creates a guiding plasma density profile moving at $v_f$. This transverse intensity structure is achieved by combining two Laguerre-Gaussian modes in orthogonal polarizations [Fig. 1(d)]. \par

When propagating in a homogeneous partially-ionized plasma, a SFF pulse designed such that $v_{g0}<v_f<c$, where $v_{g0}$ is the initial group velocity of the pulse, creates an intensity peak that moves forward within the pulse temporal envelope. The last time slice in the pulse envelope comes to focus first in time, ionizing the background medium. Photons in this initial focus self-accelerate, increasing in frequency and group velocity. These accelerated photons catch up to and remain in phase with the intensity peak formed by the un-shifted, downstream photons that are coming into focus later in time. The SFF creates a guiding, radial electron density gradient that dynamically forms just ahead of the central axial density gradient responsible for the photon acceleration. Accelerated photons, thus confined near the optical axis, overlap temporally and spatially with other photons of varying frequency. This local increase in bandwidth and photon density leads to dramatic self-steepening and elevated amplitude of the main intensity peak. The resulting sharpened axial electron density gradient causes faster frequency shifting and more sharpening, which in turn causes even faster frequency shifting. Eventually, optical wavebreaking terminates the self-steepening and limits the maximum extent of spectral broadening.\par

We discover that this novel self-shocked photon acceleration can generate fully coherent, isolated, $700 \, \mbox{as}$ pulses with continuous, multi-octave spectra extending from the $400 \, \mbox{nm}$ central wavelength of the incident pulse to $< 60 \, \mbox{nm}$ over $< 100 \, \mu \mbox{m}$ of interaction length without the need for post compression. A simple short-pass spectral filter, such as an appropriate metal-film \cite{MetalFilter}, can isolate even shorter $350 \, \mbox{as}$ pulses with ideal properties, including near-transform-limited duration and high-focusability. Under the conditions simulated, only moderate drive pulse energies of $53 \, \mu \mbox{J}$ and focused intensities of $< 5 \times 10^{16} \, \mbox{W cm}^{-2}$ were required, which would enable the use of a high-repetition rate, table-top driver with existing laser technology in an experimental realization. \par

The combination of modest drive laser requirements, relatively high conversion efficiency of $0.12 \, \%$ into $< 100 \, \mbox{nm}$ wavelengths, and the production of isolated attosecond pulses with continuous bandwidths would set a SFF-photon acceleration based source apart from both gas-based and surface-based high harmonic generation. Such sources typically produce trains of pulses and discrete bandwidths with low efficiency, unless some combination of high pulse energies (mJ-J), high intensities ($>10^{19} \, \mbox{W cm}^{-2}$) or complicated gating techniques are used \cite{Dromey2006,Sansone2011, MetalFilter}. The performance of this scheme is also in contrast to a previous proposal in which photon acceleration driven by a flying focus pulse was successfully used to limit dephasing of the upshifted photons from the ionization front, but was still limited to low efficiencies ($<10^{-3} \, \%$ ), accounting for all of the light at the accelerator output \cite{IWAVPA}. Since the spectrum inferred from photon kinetics simulations in that case showed a minimum wavelength of $>90 \, \mbox{nm}$, the efficiency to generate extreme ultraviolet light would be significantly less. The previous scheme required the use of two, couterpropagating pulses with dramatically different characteristics that needed to be carefully synchronized and injected into a pre-formed gas target, further complicating an experimental realization. Here we obtain $~ 100 \times$ the energy efficiency, over an accelerator length $~ 100 \times$ shorter, and demonstrate broadband, high intensity sub-femtosecond output pulses in a less demanding experimental configuration.
\section{Results}
\par
Self-photon acceleration of SFF pulses propagating in preionized, underdense nitrogen gas was simulated using a 2D finite-difference time-domain method. Simulation parameters used to obtain all results are shown in Table 1. Calculations were limited to the $x-z$ plane, with $\hat{z}$ the pulse propagation direction. The 3D transverse structure of the SFF is a sum of orthogonally polarized Laguerre-Gaussian modes ($LG_{l,p}$), $LG_{00} \, \hat{x} \, + \, LG_{10} \, \hat{y}$, where $l$ and $p$ are the azimuthal and radial indices respectively. $LG_{00}$ is a typical Gaussian mode, with peak intensity on the optical axis; $LG_{10}$ has zero intensity on axis which symmetrically increases to a peak off axis in the radial direction. A detailed description of the SFF pulses and their representation in 2D is included in Sec. IV. The $\hat{y}$ component (outside pulse) has higher intensity than the $\hat{x}$ component (inside pulse), such that the optical field ionization rate was high on the optical axis, but increased with absolute transverse coordinate, $|x|$, before falling to zero for large $|x|$.\par 
The SFF was spatiotemporally shaped according to the desired ionization front velocity, taking into account the nominal laser group velocity. Each time-slice within the temporal pulse envelope had a slightly different electric field radius of curvature and transverse size, such that adjacent time-slices of the pulse were focused in succession at adjacent focal positions in the far field with the same diffraction limited spot size and focused intensity. The initial and final focal lengths within the pulse envelope, and the rate of change between them, were adjusted to produce a focal spot moving at the desired velocity across the extended focal region.\par

\begin{figure}[b]
\includegraphics[scale=0.95]{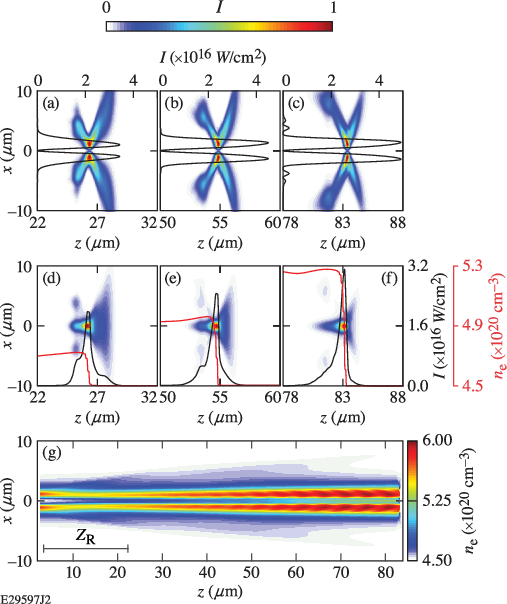}
\caption{\label{fig2}The outside pulse intensity profiles (colorbar) at (a) 96fs, (b) 192fs and (c) 288fs are shown with the corresponding transverse intensity profile (black) at the maximum intensity in each frame. The inside pulse intensity profiles (colorbar) at (d) 96 fs, (e) 192 fs and (f) 288 fs are shown with the corresponding on-axis electron density (red) and intensity (black) profiles. The color scales were normalized to the peak intensity at each time. (g) The electron density driven by the full SFF pulse after propagating for 288 fs. A guiding channel has formed over $5 Z_R$. 
}
\end{figure}
The target gas (neutral atomic density $n_0 = 10^{20} - 10^{21} \, \mbox{cm}^{-3}$) was preionized to Z=3, such that the initial electron density of the target was $n_{e0}=3 n_0$. This partial preionization serves three purposes: 1) it reduces absorption of high-frequency light, 2) it reduces the nominal laser group velocity to $v_{g0}<c$, and 3) it provides a significant level of anomalous (negative) group velocity dispersion. The ionization energies for the third and fourth valence electrons of nitrogen are $47.4 \, \mbox{eV}$ and $77.5 \, \mbox{eV}$, corresponding to wavelengths of $26 \, \mbox{nm}$ and $16 \, \mbox{nm}$. Ionization rates were calculated for $77.5 \, \mbox{eV}$, and Fourier filtering on each time step of the simulation removed wavelengths less than $26 \, \mbox{nm}$ to simulate single photon absorption by triply ionized nitrogen. The maximum ionization state allowed by the model was Z=4. Extending the model to include multi-level ionization could improve results by allowing sharper density gradients to develop and long-distance guiding to be maintained at higher intensities.
\par
\begin{table*}
\caption{\label{tab:table3}Parameters for simulations shown in each figure. $\tau$ is the total full-width at half-maximum (FWHM) duration of the pulse, $\tau_{eff}$ is the FWHM duration of the far-field intensity peak formed by the flying focus. $I_{in}$ and $I_{out}$ are the peak intensities of the inside and outside pulse respectively.  For all simulations $\lambda_0=400\mbox{nm}$, and $F/\#=6$.}
\begin{ruledtabular}
\begin{tabular}{ccccccc}
Fig.&$v_f/v_{g0}$&$\tau \, (\mbox{fs})$&$\tau_{eff} \, (\mbox{fs})$&$I_{in},I_{out} (\times 10^{16} \, \mbox{W cm}^{-2})$&$n_0 (\times 10^{20} \, \mbox{cm}^{-3})$&Input energy ($\mu \,$J)\\ \hline
2&$1.015$&12&3.3&1.2, 4.0&1.5&32\\
3&$1.015$&12&3.3&4.0, 4.8&1.5&53\\
4 (a)-(d)&$1.015$&12&3.3&4.0, 4.8&1.5&53\\
4(e)&$1.030$&30&7.3&4.0, 4.8&6.5&153\\
4(f)&NA&8.4&NA&4.0, NA&6.5&2.3\\
\end{tabular}
\end{ruledtabular}
\end{table*}

The outer pulse of the SFF forms a co-located plasma channel, dynamically guiding the inside pulse [Fig. 2]. The outside pulse [Fig. 2(a)-(c)] and the inside pulse [Fig. 2(d)-(f)] have the same spatiotemporal shaping and their temporal envelopes have zero delay. Since the outside pulse is more intense, as it comes into focus it reaches the ionization intensity threshold of the background plasma earlier in time and further along the accelerator than the inside pulse. Optical field ionization creates higher off-axis electron density, corresponding to a negative refractive index gradient in the transverse direction. Each slice of the inside pulse comes to focus just inside of the leading edge of this continuously forming optical waveguide. As a result, the guiding structure does not interfere with the propagation of the inside pulse before it comes to focus. Each slice of the inside pulse is prevented from diffracting or refracting from its self-generated plasma only after it reaches focus. This confines the slice to a  diffraction limited transverse size over the remaining length of the channel created by the outside pulse. The piece of the outside pulse that is in focus retains roughly the same shape throughout the entire extent of its propagation, such that the final guiding plasma channel extends over the entire extended focal region of the SFF—nearly five Rayleigh lengths ($Z_R= 18 \, \mu$m) [Fig. 2(g)].
\par
\begin{figure}[b]
\includegraphics[scale=1]{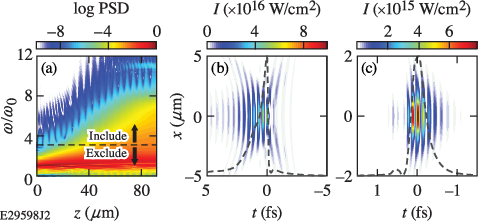}
\caption{\label{fig3} (a) Evolution of the power spectral density (PSD) on the optical axis with interaction length $z$ where $\omega_0$ is the frequency corresponding to the peak wavelength of the incident pulse, $\lambda_0 = 400 \,$nm. The black line indicates the position of the short-pass filter applied to all light passing through $z = 90 \, \mu \mbox{m}$ to obtain the pulse shown in (c). (b) The inside pulse at the output of the accelerator ($z = 90 \, \mu \mbox{m}$) with no filtering. (c) The inside pulse at $z = 90 \, \mu \mbox{m}$ after filtering out all wavelengths $ > 124 \, \mbox{nm}$.
}
\end{figure}
The inside pulse begins with high intensity at the back of the pulse envelope [Fig. 2(d)], where notably, the focal spot size was compressed to 60\% of the diffraction limit due to additional focusing by the plasma waveguide. As the high intensity light at the back of the pulse propagates with a small diameter, the on-axis intensity and electron density increase and steepen [Fig. 2(e)-(f)]. The effective on-axis duration of the pulse decreased by a factor of 1.6, from 3.3 fs to 2 fs over only $90 \, \mu$m. At higher incident intensities, the spot size was only compressed to 80-100\% of the diffraction limit, but the duration was compressed 4.7$\times$ to 700 as.
\par
The spectrum of the inside pulse evolves rapidly as it propagates along the optical axis [Fig. 3(a)]. Significant broadening about the fundamental frequency was observed within the first $15 \, \mu$m of interaction length. The spectral width at the $10^{-4}$ level more than doubled within only $45 \, \mu$m, and nearly quadrupled within $90 \, \mu$m. At this point, almost all of the light was shifted to the high-frequency side of the spectrum. Although some ionization-induced odd-harmonic generation  was observed \cite{ionization_harmonics}, comparisons with photon-kinetics simulations, which do not include nonlinear effects \cite{PhotonKinetics,PABook}, indicate that it did not play a role in the overall spectral broadening. The spectrum of the inside pulse was radially homogeneous within the diffraction limited spot ($r_0 = 1.5 \, \mu \mbox{m}$), but changed further out due to imperfect guiding at these larger radii. 
\par
\begin{figure}[b]
\includegraphics[scale=1.0]{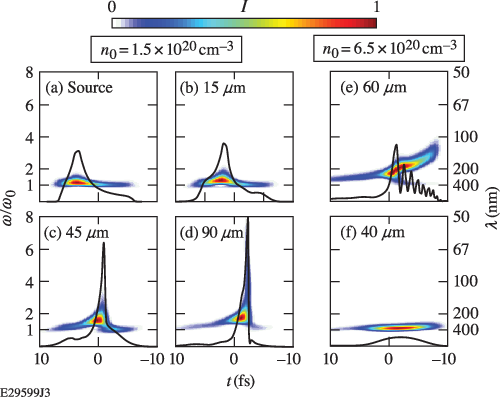}
\caption{\label{fig4} Spectrograms obtained by wavelet transform of the inside pulse electric field on the optical axis for a SFF pulse at (a) the source input plane, (b) 15 $\mu$m, (c) 45 $\mu$m, and (d) 90 $\mu$m. For comparison, the spectrograms at higher density ($n_0 = 6.5 \times 10^{20} \, \mbox{cm}^{-3}$) are shown for (e) a SFF pulse with $v_f = 1.030 \, v_{g0} = 0.8740 \, c$ and (f) a standard Gaussian pulse. The color scale in each plot is normalized to the peak intensity. On-axis intensity profiles (black) are normalized to $10^{17} \, \mbox{W cm}^{-2}$.
}
\end{figure}
At high intensities the total duration of the pulse was compressed considerably during the interaction, such that an isolated $700 \, \mbox{as} $ pulse was obtained at the accelerator output ($z = 90 \, \mu$m) [Fig. 3(b)]. This pulse could be useful due to its short duration, high intensity and broad bandwidth, however it has complicated spatial, spectral, and temporal correlations that could make it unsuitable for certain applications. Applying a simple short-pass filter to the output of the photon accelerator isolates a spatially coherent, extreme ultraviolet pulse [Fig. 3(c)]. The pulse, containing only wavelengths shorter than $124 \, \mbox{nm}$ and 0.21\% of the total incident pulse energy has a duration of 1.5 optical cycles, or $350 \, \mbox{as}$ intensity full-width at half-maximum --- 1.3 times the transform limit for the filtered bandwidth. With a nearly flat quadratic phase in the transverse direction, the pulse has ideal focusability. Spectral filtering could be achieved by allowing the output pulse to naturally diffract out of the end of the accelerator, collimating it with an appropriate curved reflector \cite{BBReflect}, and then allowing it to pass through a $200 \, \mbox{nm}$ thick magnesium foil \cite{MetalFilter}. Finally the pulse could be refocused through another curved reflector for use in an experiment. The outside pulse undergoes less spectral broadening than the inside pulse, but there is still significant spectral overlap between them. For longer wavelength spectral filter cutoffs ($\gtrsim 150 \, \mbox{nm}$) a pinhole spatial filter or polarization filter could be used to remove the outside pulse. \par
The formation of an optical shock drives the extreme refractive index gradients that are responsible for the rapid spectral broadening and temporal compression of the pulse [Fig. 4]. The intensity of the incident pulse peaks at the back of the pulse envelope where the first focus occurs [Fig. 4(a)], and the electron density gradient created by this peak induces a frequency up-shift of the photons in the initial focus. These photons experience an increase in group velocity due to the anomalous (negative) group velocity dispersion in a plasma, \emph{i.e.}, $v_g = c (1-\omega_p^2/\omega^2)^{1/2}$. Higher frequencies, initially generated at the back of the pulse, catch up to the un-shifted downstream photons, causing a local increase in the intensity and bandwidth, which creates the observed self-steepening and temporal compression. 
\par
After 15$ \, \mu$m [Fig. 4(b)], the bandwidth temporally overlapped with the intensity peak at the fundamental frequency has increased. The peak intensity has steepened on the leading edge and moved forward within the pulse envelope. By 45$ \, \mu$m [Fig. 4(c)], significant local spectral broadening and self-steepening has doubled the peak intensity. At 90$ \, \mu$m [Fig. 4(d)], a temporally-localized, extremely broadband signal has developed. Notably the intensity increases from near zero to its peak in half an optical cycle, indicating that a strong optical shock has formed on the leading edge of the pulse. This time-localized spectral broadening is a unique feature of the self-shocked photon acceleration regime that facilitates attosecond pulse generation without any post-accelerator phase compensation.
\par
For an initial density 4.3$\times$ larger than in the previously discussed simulations, the shock catastrophically collapsed over less than $60 \, \mu \mbox{m}$ [Fig. 4(e)]. The spectrogram can be considered a time-velocity phase space representation, and the observed ``folding" of the spectrogram, such that non-adjacent frequencies (velocities) overlap in time indicates optical wavebreaking \cite{OWBTheory, OWBExp}. A strongly modulated intensity profile developed on the leading edge of the pulse, and distinct peaks emerged in the spectrum due to nonlinear four wave mixing. These two effects terminate the self-steepening and spectral broadening of the main intensity peak by extracting energy and causing deleterious plasma formation.
\par
Figure 4(f) demonstrates the limitations of a standard 8.4 fs Gaussian pulse that was focused at $z = 10 \, \mu$m. After $40 \, \mu$m, the leading edge of the pulse shifted from 400 nm to 370 nm, but the trailing edge completely diverged from the optical axis due to plasma-refraction and diffraction. Refraction and diffraction reduced the peak intensity by $> 4\times$, such that ionization and photon acceleration ceased.
\par
Self-steepening of Gaussian pulses was observed in simulations, but the effect was weak compared to the SFF pulse. As photons frequency shifted at the peak of the Gaussian envelope, their increase in group velocity caused them to temporally overlap with the lower intensity time-slices ahead, which were already diffracting out of the focus. This resulted in a small increase in local bandwidth, but no increase in the intensity, such that self-steepening was negligible. For the SFF pulse, photons accelerated in the initial focus were immediately injected into a new high-intensity focus. Each subsequent time slice that came to focus had the original intensity and bandwidth of the pulse, in addition to the accelerated photons, leading to a coherent buildup of intensity and bandwidth. The accelerating gradient was continually pumped by a higher, sharper intensity peak until it collapsed due to wavebreaking.
\par
\begin{figure}[b]
\includegraphics[scale=1.0]{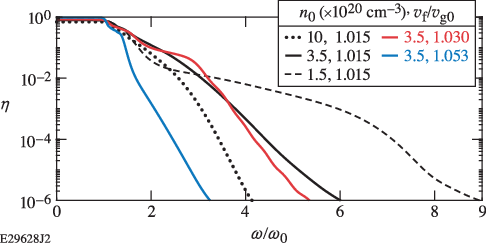}
\caption{\label{Fig5} Parameter scaling of focal velocity $v_f$ and initial density $n_0$ for a $90 \, \mu$m accelerator. $\eta$ is the energy efficiency to produce all light at frequencies above the frequency indicated on the horizontal axis. For all data shown $I_{in} = 4.0 \times 10^{16} \, \mbox{W cm}^{-2} $ and $I_{out} = 4.8 \times 10^{16} \, \mbox{W cm}^{-2} $.}.
\end{figure}
\section{Discussion and Conclusion}
The exact combination of density and flying focus velocity must be optimized to deliver the greatest bandwidth increase [Fig. 5]. Clearly, $v_f \gtrsim v_{g0}$ is optimal for all densities. For $v_f < v_{g0}$, photons focused to shorter focal lengths are contained in earlier time slices of the pulse envelope, and plasma formation prevents the ionization front from propagating continuously over a long distance \cite{IWAVTheory, IWAVExp, IWAVExp2}. For $v_f \gg v_{g0}$, photons up-shifted at the start of the accelerator do not obtain a large enough group velocity to catch up to the main intensity peak, so a strong shock cannot form.
\par
A target with higher initial density has the potential to reach a higher maximum electron density, in principle supporting sharper density gradients and faster frequency shifting. However, a correspondingly larger initial electron density also yields greater group velocity dispersion and a smaller nominal group velocity, which leads to a shorter wave-breaking distance. At large initial electron density, for the observed optimum $v_f \gtrsim v_{g0}$, the initial intensity peak forms a sharp density gradient, and dispersion is strong, quickly accelerating photons to $v_g \gg v_f$. A shock forms and collapses over a short distance since even a small increase in frequency results in a large increase in group velocity, and the initial $v_f \ll c$ is easy to overtake for a photon of higher frequency. 
\par
At smaller initial densities, a shallower initial density gradient formed, and dispersion was weak, so the shock was formed and sustained over a longer distance. Here a large increase in frequency resulted in a much smaller increase in group velocity, and since $v_f \lesssim c$, the shifted photons took longer to catch up to, and then dephase from the intensity peak at the fundamental frequency. The low initial density case leads to larger overall spectral broadening because the shock can build up a density profile with a sharper scale length, sustain its propagation over a longer distance and remain in phase with accelerated photons over a longer distance than in the high density case. Notably, the spectrum was less sensitive to the exact $v_f$ at higher densities, since initial frequency shifts were larger and dispersion was more significant.
\par
Experimentally, a separate laser pulse could be used to preionize the target through optical field ionization, or alternatively the plasma could be heated to achieve the desired average ionization state, $Z$. For the $90 \, \mu \mbox{m}$ accelerator presented here, a longer plasma could be generated, with the interaction only occurring in the first section. Excess energy used in ionizing the unused plasma would decrease the overall energy efficiency by a small factor. An optimal accelerator would require plasmas longer than $90 \, \mu \mbox{m}$, since lower densities and longer acceleration lengths are required for larger frequency shifts.  Although small density fluctuations in the target would not appreciably affect the interaction, larger fluctuations could affect the focusing of the pulse leading to a degradation of the guiding channel driven by the outside beam, and ultimately to a lower quality output pulse.

\par
In summary, a new approach to coherent, broadband radiation generation in the extreme ultraviolet, enabled by the combination of state-of-the-art pulse shaping methods, was proposed and demonstrated computationally. Optical shock-enhanced self-photon acceleration resulted in coherent spectral broadening to extreme ultraviolet frequencies over a fraction of the distance required in a typical Kerr-based supercontinuum source. Self-compression of the incident pulse was significant, and ideal attosecond pulses could be isolated with a simple short-pass filter. Such a pulse could be useful for enabling a single-shot version of the coherence tomography application discussed in Ref. \cite{XCT_multishot1} and providing sub-fs time resolution to observe transient effects. A photon accelerator based on SFF driven plasma could be optimized further to deliver higher frequencies, shorter pulses and better efficiency. By tailoring the initial axial density profile of the target gas and the spatiotemporal shaping together, a shock could be generated at higher densities, accelerating and then stabilizing over long-distances at lower densities \cite{PAAccel,FFVacuumAccel}, which could enable the generation of shorter extreme ultraviolet wavelengths, and possibly a table top-source of coherent soft x-rays.
\par

\section{Appendix: SFF field model}

A novel field injection scheme was used to simulate structured flying focus pulses. The desired field amplitude to be injected on each time step is calculated analytically using well known equations of Gaussian optics and simple algebraic equations that govern the desired spatiotemporal and transverse intensity profile characteristics. This calculation method amounts to injecting a Laguerre-Gaussian beam at a single axial position that has time varying focal length, but constant $F/\#$. This model captures the salient features of a flying focus-like pulse, without reference to a particular spatiotemporal shaping method \cite{FlyingFocus1,FlyingFocus2,DLWFA,DLWFA2,NLFF,Abouraddy}. \par

The field is injected at an axial distance of $z = 0$. Different time slices in the injected pulse are focused to different axial positions, where the shortest focal position is defined as $z = f_s$ and the longest as $z = f_l$. The length of the ``extended focal region," which defines the length over which the peak intensity of the pulse will propagate in the far field, is therefore $l = f_l - f_s$. For a preionized target the initial group velocity is calculated as $v_{g0} = c \, \sqrt{1-n_e/n_c(\lambda_0)}$, where $n_e$ is the initial electron density of the target, $n_c(\lambda_0)$ is the critical plasma density for the initial, central wavelength of the injected field, $\lambda_0$, and $c$ is the vacuum speed of light.\par

The field is injected in vacuum just outside the edge of the medium, so that the free space propagation distance is negligible, and the first temporal slice of the pulse is injected at time $t = 0$. The time for the field to propagate from the injection point to $z=f_s$ is $t_s=f_s/v_{g0}$. An inherent delay between field propagating to $f_s$ and $f_l$ within the pulse is denoted $\Delta t$, which can take either positive or negative sign, such that the field focusing at $f_l$ arrives there at $t_l = f_l/v_{g0} + \Delta t$. For a constant focal velocity
\begin{equation}
    v_f =  \frac{dz}{dt} = \frac{f_l-f_s}{t_l-t_s} = \frac{l}{\frac{l}{v_{g0}} + \Delta t} ~ .
\end{equation}
Solving for $\Delta t$ gives the time required to sweep between focal lengths $f_s$ and $f_l$,
\begin{equation}
    \Delta t = l\frac{v_{g0}-v_f}{v_{g0} v_f} ~ .
\end{equation}
Since $v_{g0}$ and $l$ are positive, the sign of $\Delta t$ depends only on the choice of $v_f$. Defining $f_0$ as the focal length injected at $t=0$,

\begin{equation}
    f_0 = 
    \begin{cases}
        f_l ~~~~~ \text{if} ~~ \Delta t < 0  ~~~ (v_f < 0 ~~ \textrm{or} ~~ v_f > v_g) \\
        f_s ~~~~~ \text{if} ~~ 0 < \Delta t  ~~~ (0 < v_f < v_g),
    \end{cases}
\end{equation}
the focal length to be injected as a function of time is given by
 
 \begin{equation}
     f(t) = f_0 + \frac{l}{\Delta t} t ~ .
 \end{equation}

\par

The function $f(t)$ can be used to calculate the full field to be injected at every time step of the simulation for any field that is analytically expressable in terms of the propagation distance, $z$. Here we limit the calculation to the family of Laguerre-Gaussian modes, $LG_{l,p}$, where $l$ and $p$ are the azimuthal and radial indices respectively. For all $LG_{l,p}$ the far field beam waist radius and the Rayleigh length are given by $w_0 = 2 \lambda_0 (F/\#) /\pi$ and $Z_R = \pi w_0^2/\lambda_0$ respectively. We substitute $f(t)$ for the general focal length $f$ in the well know equations for the field curvature $R(t) = f(t)[1+\left(Z_R/f(t)\right)^2]$,
beam radius $W(t) = w_0[1+\left(f(t)/Z_R\right)^2]^{1/2}$, and Gouy phase $\psi(t) = \textrm{tan}^{-1}\left(f(t)/Z_R\right)$ such that they are all parametric functions of time. \par
 
 The full field to be injected as a function of time and transverse coordinates $r$ and $\theta$ becomes,
    \begin{multline}
     E(r,\theta,t) = E_0\frac{w_0}{W(t)} L^{|l|}_p \left[\frac{2 r^2}{W^2(t)}\right] \\
     \times ~ \textrm{exp} \left[ -\frac{r^2}{W^2(t)} \right] \textrm{exp} \left\{ -i \left[ \omega t - k \frac{r^2}{2R(t)} \right]\right\}\\
     \times ~ \textrm{exp} \left\{ i \left[   l\theta - (1 + |l| + p) \psi(t) \right] \right\} T(t)  ,
    \end{multline}
 where $L^{|l|}_p$ is the generalized Laguerre polynomial of order $|l|$ and $p$, $E_0$ is the electric field amplitude at focus, and $\omega$ and $k$ are the vacuum frequency and wavenumber for $\lambda_0$. $T(t)$ is a time envelope function, which was chosen to be a super-Gaussian of order $g=8$, centered at $t=t_0$ and with width $\tau$,
 
 \begin{equation}
     T(t) = \textrm{exp} \left\{ - \left[ \frac{(t-t_0)^2}{2 \tau^2} \right]^g \right\} ~ .
 \end{equation}
 where $t_0=|\Delta t|/2$ and $\tau=\sqrt{8^{-1} \, \textnormal{log}(2)^{-1/g} \, | \Delta t|^2}$. The total energy of the pulse scales with the total pulse duration as $P_0 | \Delta t |$ where $P_0$ is the instantaneous power, such that the incident energy depends on $l$, $v_f$ and $v_{g0}$. The duration of the intensity peak in the focal region is $\tau_{eff}=2 Z_R/|U|$ where $U=v_g v_f/(v_g-v_f)$, which in 2D becomes $\tau_{eff}=2 \sqrt{3} Z_R/|U|$. 
 \par
 
 For 2D simulations we set $y=0$ in Eq. (5), and take the prefactor $w_0/W(t) \rightarrow \sqrt{w_0/W(t)}$. The spatiotemporal characteristics defined by $f(t)$ were the same for the two polarization states. A mode with $l=p=0$ was injected with $\hat{x}$ polarization, and a mode with $l=1$ and  $p=0$ was injected with orthogonal $\hat{y}$ polarization. Since Laguerre-Gaussian modes of unequal order experience a different Gouy phase shift, injecting them in the same polarization yields a transverse intensity profile that evolves with $z$. Therefore to maintain the desired guiding-intensity profile for all $z$, the two modes are injected with orthogonal polarizations so that no interference occurs between them.\par
 
 In practice such a pulse could be produced by splitting the Gaussian output of a master oscillator, rotating the polarization of one arm, allowing the other arm to pass through a spiral phase plate \cite{SpiralPP1,OAMReview,OASpiral}, then recombining the two arms such that the relative timing between the centers of the two pulse envelopes is zero. Such splitting and recombining of a pulse from the same laser oscillator can be achieved with small temporal uncertainty, as in a typical Mach-Zehnder interferometer. Small deviations from this co-timing condition are not expected to significantly impact the results. Finally, the recombined pulse could be spatiotemporally shaped using one of the several methods discussed in references \cite{FlyingFocus1,FlyingFocus2,DLWFA,DLWFA2,NLFF,Abouraddy}.\par
 
\begin{acknowledgments}
This material is based upon work supported by the Office of Fusion Energy Sciences under Award Number DE-SC0019135, the Department of Energy National Nuclear Security Administration under Award Number DE-NA0003856, the University of Rochester, and the New York State Energy Research and Development Authority.\par 
This report was prepared as an account of work sponsored by an agency of the U.S. Government. Neither
the U.S. Government nor any agency thereof, nor any of their employees, makes any warranty, express or implied,
or assumes any legal liability or responsibility for the accuracy, completeness, or usefulness of any information,
apparatus, product, or process disclosed, or represents that its use would not infringe privately owned rights.
Reference herein to any specific commercial product, process, or service by trade name, trademark, manufacturer,
or otherwise does not necessarily constitute or imply its endorsement, recommendation, or favoring by the U.S.
Government or any agency thereof. The views and opinions of authors expressed herein do not necessarily state
or reflect those of the U.S. Government or any agency thereof. \par
The authors would like to thank D. Turnbull and J. L. Shaw for fruitful discussions and support.
\end{acknowledgments}

\bibliography{osespa}

\end{document}